\documentclass[aps,twocolumn,superscriptaddress]{revtex4-2}

\usepackage[T1]{fontenc}
\usepackage{amssymb}
\usepackage{amsthm}
\usepackage{xfrac}
\usepackage{graphicx}
\usepackage{amsmath}
\usepackage{amsfonts}
\usepackage{color}
\usepackage{centernot}
\usepackage{mathtools}
\usepackage{stmaryrd}
\usepackage{xcolor}
\usepackage{url}

\begin{document}

\title{Why is the forward cascade intermittent but the inverse cascade not?}
\author{Alexei A. Mailybaev}
\affiliation{Instituto de Matem\'atica Pura e Aplicada -- IMPA, Rio de Janeiro, Brazil}
\email{alexei@impa.br}

\begin{abstract}
Forward and inverse energy cascades exhibit
a striking contrast in intermittency,
but the mechanism underlying this asymmetry
remains poorly understood.
Using a stochastic shell model, we develop
a first-principles perturbative theory of
cascade reversal.
We show that the direction of energy transfer
controls the growth or decay of fluctuations
across scales,
selecting Kolmogorov (K41) statistical symmetry
in the inverse cascade and a weaker hidden
symmetry of shell multipliers in the forward cascade.
This change of symmetry produces a singular
statistical transition reminiscent of a
mixed-order phase transition: the 
anomalous corrections jump at cascade reversal,
while the correlation length diverges.
Our results identify a dynamical mechanism
linking cascade direction, statistical
symmetry breaking, and intermittency.
\end{abstract}

\date{\today}
\maketitle

The direction of turbulent energy transfer can change
under confinement, rotation, or stratification,
giving rise to forward, inverse, and coexisting
cascades, sometimes separated by sharp transitions
\cite{smith1996crossover,celani2010turbulence,
benavides2017critical,sahoo2017discontinuous,
vankan2020critical,alexakis2024atmospheres,
alexakis2026anisotropic,gome2025helicity}.
These regimes also differ in their statistical
properties: the forward energy cascade in
three-dimensional turbulence is intermittent,
whereas the inverse energy cascade in two
dimensions is nearly self-similar
\cite{boffetta2000inverse,boffetta2012twoD,
alexakis2018cascades}.
Previous studies have offered physical and
phenomenological arguments for this contrast,
emphasizing quasi-equilibrium statistics
in inverse cascades
\cite{lvov2002quasigaussian,gilbert2002inverse}
and the role of interscale energy-transfer times
\cite{benzi2005dynamical}.
Cascade models and statistical descriptions
based on interscale multipliers have also
provided insights into anomalous scaling
\cite{benzi1993intermittency,eyink2003gibbsian,
chen2003kolmogorov}.
Despite these advances, a first-principles
explanation linking cascade reversal,
statistical scaling symmetries, and
intermittency remains elusive.

Here we investigate this connection using a
stochastic shell model combining the
Desnyansky--Novikov (DN) and Obukhov interactions,
previously studied in deterministic mixed
cascade models \cite{dombre1998intermittency}.
The relative strengths of the two interactions
determine the direction of the K41 energy flux,
while an independent parameter controls the
stochastic transfer fluctuations.
The stochastic extension preserves energy
conservation and the scaling symmetries of
the deterministic dynamics, allowing a systematic
perturbative analysis around the K41 state. 
Building on the hidden scaling symmetry
of turbulence
\cite{mailybaev2021hidden,mailybaev2022shell}
and its recent perturbative developments
\cite{mailybaev2026perturbative,mailybaev2026first},
we derive the statistics of
both cascade regimes and the singular transition
between them from the ideal equations of motion.
We show that the direction of energy transfer
controls the growth or decay of fluctuations
across scales, determining which statistical
scaling symmetry survives and explaining
the discontinuous onset of intermittency
at cascade reversal.

\vspace{2mm}
\textit{Mixed shell model.}
Shell models use a geometric sequence of wavenumbers
$k_n=\lambda^n$, where $\lambda>1$ is the shell-spacing
parameter (typically $\lambda=2$)
\cite{biferale2003shell}.
Each shell is associated with a real-valued amplitude
$u_n(t)$, representing the characteristic velocity
fluctuation at scale $\ell_n=k_n^{-1}$.
We first introduce the deterministic ideal model,
\begin{equation}
    \frac{du_n}{dt}
    =u_{n-1}F_{n-1}-u_{n+1}F_n,
    \label{eq:mixed_model}
\end{equation}
with transition rates
\begin{equation}
    F_n=k_n\left[
        (1+a)\lambda^{1/3}u_n
        -(1-a)\lambda^{2/3}u_{n+1}
    \right].
    \label{eq:det_transition_rates}
\end{equation}
The parameter $-1\leq a\leq1$ interpolates
between the pure DN model at $a=1$
and the pure Obukhov model at $a=-1$ \cite{desnyansky1974evolution,obukhov1971some}.

The model possesses scaling symmetries
\begin{equation}
    u_n(t)\mapsto
    \lambda^h u_{n+1}\left(\lambda^{h-1}t\right),
    \label{eq:holder_scaling}
\end{equation}
where $h\in\mathbb R$ is an arbitrary
H\"older exponent. 
It also admits the Kolmogorov (K41)
stationary solution
\begin{equation}
    u_n^*=k_n^{-1/3}=\lambda^{-n/3}.
    \label{eq:K41}
\end{equation}
More generally, any multiple of this
solution is also stationary.
The K41 solution \eqref{eq:K41} is
invariant under the scaling symmetry
\eqref{eq:holder_scaling} for $h=1/3$.

The energy balance for each shell
takes the form
\begin{equation}
    \frac{d}{dt}\frac{u_n^2}{2}
    =\Pi_{n-1}-\Pi_n,
    \label{eq:energy_balance}
\end{equation}
where the energy flux through shell $n$ is $\Pi_n=u_nu_{n+1}F_n$.
For the stationary K41 solution \eqref{eq:K41},
the flux is independent of $n$ and equals $\Pi_*=2a$.
Thus, the K41 solution 
corresponds to a forward energy cascade
when $a>0$ and to an inverse energy
cascade when $a<0$.

We linearize the dynamics around the K41 solution by introducing
\begin{equation}
    u_n(t)=k_n^{-1/3}
    \left[1+\varepsilon v_n(t)\right],
    \label{eq:perturbation}
\end{equation}
where $\varepsilon$ is a small parameter
and $v_n$ describe relative fluctuations.
Substituting into \eqref{eq:mixed_model}
and retaining first-order terms
yields the discrete advection--diffusion equation
\begin{equation}
    \rho^{-n}\frac{dv_n}{dt}
    +3\Pi_*
    \frac{v_{n+1}-v_{n-1}}{2}
    =v_{n+1}-2v_n+v_{n-1},
    \label{eq:linear_v}
\end{equation}
where $\rho=\lambda^{2/3}>1$.
This equation represents the linearized
shell-energy balance, since
$u_n^2/2=\rho^{-n}/2+\varepsilon\rho^{-n}v_n
+O(\varepsilon^2)$.
The transport coefficient $3\Pi_*$ reflects
the cubic dependence of the energy flux
on velocity amplitudes.
Thus, energy fluctuations are transported
in the direction of the cascade, determined
by the sign of the flux $\Pi_*=2a$,
while the diffusive term on the right-hand
side smooths their variations across scales.

Seeking stationary perturbations
$v_n=r^n$ yields the characteristic roots
\begin{equation}
    r_1=1,
    \qquad
    r_2=(1+3a)/(1-3a).
    \label{eq:stationary_eigenvalues}
\end{equation}
The first represents a uniform perturbation
of the K41 amplitude, while the second
preserves the energy flux to first order.
For this flux-preserving mode, diffusion
must compensate the transport induced by
the energy cascade.
Since diffusion acts against amplitude
gradients, this balance requires the mode
to grow in the direction of energy transfer:
$|r_2|>1$ in the forward cascade ($a>0$),
whereas $|r_2|<1$ in the inverse case ($a<0$).

\vspace{2mm}\textit{Stochastic transfer rates.}
In the spirit of Kraichnan-type models
\cite{falkovich2001particles,wirth1996anomalous,
benzi1997analytic},
we introduce Gaussian white-in-time
driving into the energy transfer
between neighboring shells, while retaining
its nonlinear and conservative structure.
Specifically, we define
\begin{equation}
    \begin{aligned}
    dF_n={}&
    k_n\left[
        (1+a)\lambda^{1/3}u_n
        -(1-a)\lambda^{2/3}u_{n+1}
    \right] dt\\
    &+\varepsilon(k_n|u_n|)^{1/2}
    \circ dW_n.
    \end{aligned}
    \label{eq:stochastic_transition_rates}
\end{equation}
Here $\varepsilon\geq0$ controls the amplitude
of stochastic fluctuations, and $W_n$ are
independent standard Brownian motions.
The stochastic model is obtained from
\eqref{eq:mixed_model} by replacing
$F_n\,dt$ with $dF_n$, with all products
interpreted in the Stratonovich sense.
The square-root dependence of the stochastic
transfer rates in
\eqref{eq:stochastic_transition_rates}
ensures that the scaling symmetries
\eqref{eq:holder_scaling} remain valid in law.
The Stratonovich formulation preserves the local energy
balance \eqref{eq:energy_balance}, with $\Pi_n\,dt$
replaced by the stochastic flux
$d\mathcal J_n=u_nu_{n+1}\circ dF_n$.

In the stochastic formulation, the leading-order
fluctuations \eqref{eq:perturbation} obey
the stochastic counterpart of the
advection--diffusion equation \eqref{eq:linear_v},
\begin{equation}
    \begin{aligned}
    \rho^{-n}dv_n
    &+3\Pi_*\frac{v_{n+1}-v_{n-1}}{2}\,dt\\
    &=(v_{n+1}-2v_n+v_{n-1})\,dt\\
    &\quad+\rho^{-n/2}
    \big[dW_{n-1}-\rho^{-1/2}dW_n\big],
    \end{aligned}
    \label{eq:OU_system}
\end{equation}
which is an infinite-dimensional
Ornstein--Uhlenbeck (OU) system.
As we show below, the growth or decay
of the flux-preserving mode $r_2$ determines
whether this system admits 
K41-invariant stationary covariances,
thereby linking the direction of energy
transfer to statistical scaling symmetry.

\vspace{2mm}
\textit{K41 inverse cascade solution.}
We seek a centered stationary Gaussian measure
for the OU system \eqref{eq:OU_system},
invariant under the K41 scaling symmetry.
Its covariance is
$C_{ij}=\langle v_i v_j\rangle_0$,
where $\langle\cdot\rangle_0$ denotes
averaging with respect to the Gaussian measure.
For the perturbation variables \eqref{eq:perturbation},
the K41 symmetry
\eqref{eq:holder_scaling} with $h=1/3$ takes the form
\begin{equation}
    v_n(t)\mapsto v_{n+1}(\rho^{-1}t).
    \label{eq:K41_symmetry_v}
\end{equation}
Hence, the covariance must be Toeplitz,
$C_{ij}=c_{i-j}$, with $c_{-j}=c_j$.
The Lyapunov equation for the covariance matrix then reduces to (see SM~\cite[\S I\,A]{SM})
\begin{equation}
    \begin{aligned}
    &3\Pi_*\frac{\rho^j-1}{\rho^j+1}
    \frac{c_{j+1}-c_{j-1}}{2} = c_{j+1}-2c_j+c_{j-1}\\
    &\qquad\qquad
    +\frac{(1+\rho^{-1})\delta_{j0}
    -\delta_{j1}}{\rho^j+1}, \qquad j\geq0.
    \end{aligned}
    \label{eq:c_recurrence}
\end{equation}
For large shell separations $j\gg1$,
the covariance balance is dominated by
the faster small-scale dynamics and reduces
to the stationary form of
Eq.~\eqref{eq:linear_v}.
Hence, the asymptotic behavior of $c_j$
is governed by the same flux-preserving
mode $r_2$. 
A bounded Toeplitz solution requires
this mode to decay with shell separation.
This is possible in the inverse cascade,
$|r_2|<1$, but not in the forward cascade,
$|r_2|>1$.

In the inverse-cascade regime $a<0$,
the Lyapunov equation \eqref{eq:c_recurrence}
admits an explicit family of solutions
(see SM~\cite[\S I\,B]{SM}). 
The solution is uniquely determined up to
an additive constant
$c_\infty=\lim_{j\to\infty}c_j$,
corresponding to common fluctuations of
the K41 amplitude and hence of the energy flux.
The remaining correlations $c_j-c_\infty$
decay with separation, ensuring locality
in scale.
The K41-invariant statistics imply
nonintermittent scaling of the time-averaged structure
functions $S_p(n)=\langle|u_n|^p\rangle_t
\propto k_n^{-\zeta_p}$, with $\zeta_p=p/3$.

\begin{figure}
\centering
\includegraphics[width=0.45\textwidth]{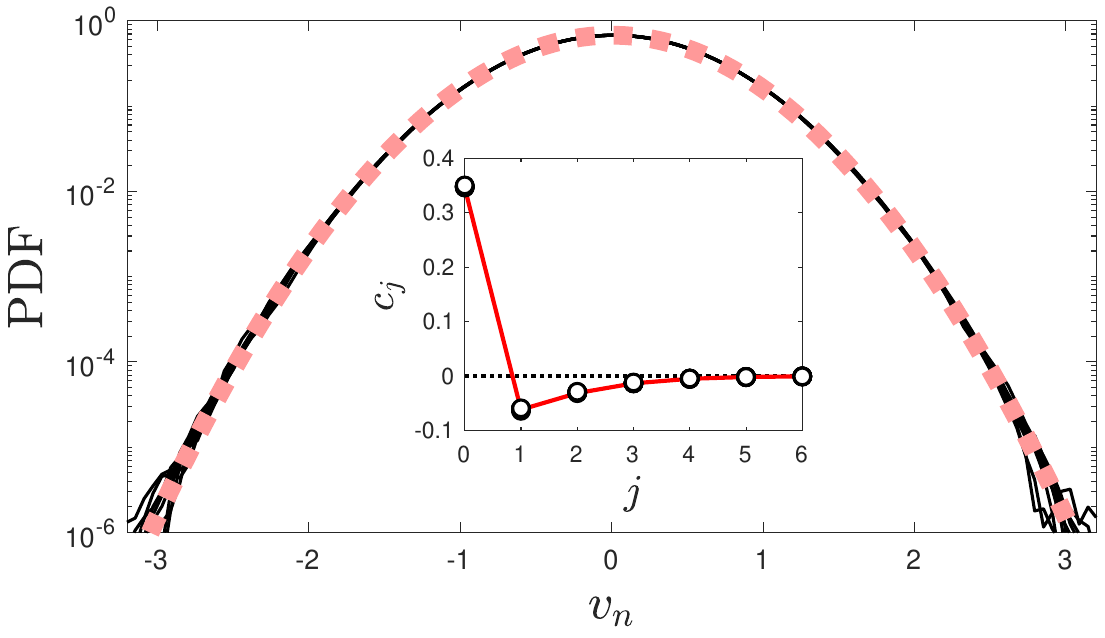}
\caption{
Gaussian K41 statistics in the inertial range
of the inverse cascade for $a=-0.2$,
$\varepsilon=0.1$, and an ultraviolet cutoff
at $N=22$. 
Probability densities of normalized shell
perturbations $v_n$ for $n=6,\ldots,14$
(black curves), compared with the theoretical
Gaussian distribution with variance
$c_0=0.3459$ (red dotted curve).
Inset: numerical covariances $\operatorname{Cov}(v_n,v_{n+j})$ as functions of
shell separation $j$ (circles), compared
with the analytical values $c_j$ for
$c_\infty=0$ (red line).
}
\label{fig1}
\end{figure}

We compare our analytical results with
long-time numerical simulations of the full
nonlinear stochastic model with infrared
and ultraviolet cutoffs.
For the forward cascade, forcing acts at large
scales and dissipation at small scales; for the
inverse cascade, their roles are reversed
(see SM~\cite[\S IV]{SM} for details).
In the inverse-cascade regime, simulations
yield a negative energy flux close to
its K41 value $\Pi_*=2a$ for all tested $a<0$.
Figure~\ref{fig1} presents results for $a=-0.2$.
The probability densities of the normalized
perturbations $v_n=(k_n^{1/3}u_n-1)/\varepsilon$
agree with the analytical Gaussian distribution.
The inset shows similarly good agreement
between the numerical covariances
$\operatorname{Cov}(v_n,v_{n+j})$ and
their analytical values.

\vspace{2mm}\textit{Intermittent forward cascade.}
For $a>0$, the covariance recurrence
\eqref{eq:c_recurrence} yields unbounded $c_j$
as $j\to\infty$, reflecting the growth factor
$|r_2|>1$.
Since covariance coefficients must satisfy
$|c_j|\leq c_0$, no K41-invariant stationary
Gaussian state can exist.
However, a weaker statistical self-similarity
survives for the shell multipliers
\begin{equation}
    \frac{u_{n+1}}{u_n}
    =\lambda^{-1/3}
    \left[1+\varepsilon x_n+O(\varepsilon^2)\right],
    \label{eq:multiplier_linear}
\end{equation}
where $x_n=v_{n+1}-v_n$.
Subtracting Eq.~\eqref{eq:OU_system}
at indices $n+1$ and $n$ yields
a closed OU system,
\begin{equation}
    \begin{aligned}
    &dx_n=(\Phi_{n-1}-\Phi_n)\,dt\\
    & -\rho^{n/2}
    \big[dW_{n-1}
    -(\rho^{1/2}+\rho^{-1/2})dW_n
    +dW_{n+1}\big],
    \end{aligned}
    \label{eq:OU_x}
\end{equation}
where $\Phi_n$ is the flux of multiplier fluctuations,
\begin{equation}
    \Phi_n=\rho^{n+1}\left[
    \frac{3\Pi_*}{2}(x_n+x_{n+1})
    -(x_{n+1}-x_n)\right].
\end{equation}
The deterministic part admits stationary
spatial modes $x_n=r^n$ with constant
$\Phi_n$, yielding two roots:
$r'_1=\rho^{-1}<1$ and $r'_2=r_2$.
Thus, the marginal mode $r_1=1$ of the
velocity fluctuations is replaced by a
decaying mode carrying a constant multiplier
flux, while the original flux-preserving
mode $r_2$ remains unchanged.

At leading order, the hidden scaling symmetry
\cite{mailybaev2021hidden,mailybaev2026perturbative,
mailybaev2026first}
acts on the multiplier fluctuations as
\begin{equation}
    x_n(t)\mapsto x_{n+1}(\rho^{-1}t),
\end{equation}
leaving the OU system \eqref{eq:OU_x}
invariant in law.
We therefore seek a stationary Gaussian state
with Toeplitz covariance
$\langle x_i x_j\rangle_0=\kappa_{i-j}$,
where $\kappa_{-j}=\kappa_j$.
For both cascade directions, the corresponding
Lyapunov equation admits a Toeplitz
solution with decaying correlations
(see SM~\cite[\S II]{SM}).
In the forward cascade, the mode $r'_1$
governs the decay $\kappa_j\propto\rho^{-j}$
at large shell separations.
Thus, although the velocity fluctuations $v_n$
do not admit a K41-invariant stationary state
in this regime, their differences $x_n$
retain scale-invariant Gaussian statistics.

The stationary Gaussian statistics of the multiplier
fluctuations $x_n$ determine the leading anomalous
corrections to the structure-function exponents,
which quantify intermittency in the forward cascade.
Applying the perturbative hidden-symmetry theory
\cite{mailybaev2026perturbative,mailybaev2026first}
to the mixed model, we obtain (see SM~\cite[\S III]{SM})
\begin{equation}
    \zeta_p=\frac{p}{3}
    -\frac{\varepsilon^2p(p-3)}
          {2\log\lambda}\mathcal K
    +O(\varepsilon^4),
    \quad
    \mathcal K=\sum_{j\in\mathbb Z}\kappa_j.
    \label{eq:scaling_exponents}
\end{equation}
For $a>0$, we have $\mathcal K>0$, implying
positive anomalous corrections for $0<p<3$
and negative ones for $p>3$.
At $p=3$, the correction vanishes, consistently
with the constant energy flux.
In contrast, for $a<0$, the covariance yields
$\mathcal K=0$, in agreement with the K41
self-similarity established for the inverse cascade.

\begin{figure*}
    \centering
    \includegraphics[width=0.87\textwidth]{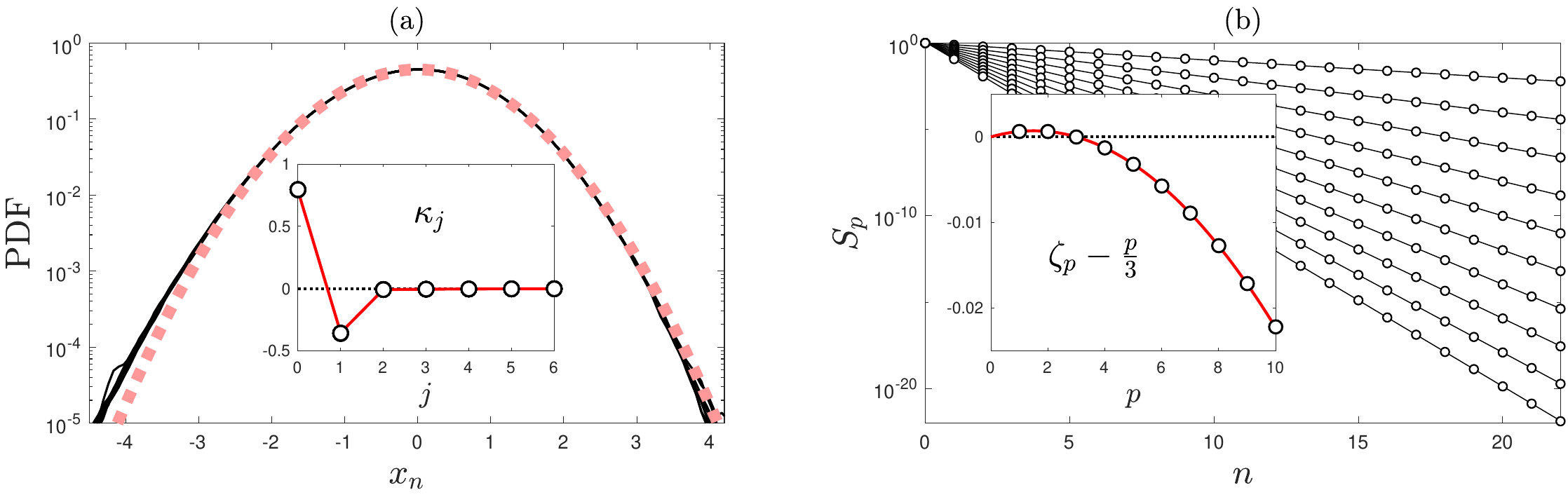}
\caption{
Multiplier self-similarity and intermittency
in the inertial range of the forward cascade
for $a=0.2$, $\varepsilon=0.1$, and $N=22$.
(a) PDFs of logarithmic multiplier fluctuations
$x_n^{\mathrm{log}}$ for $n=6,\ldots,16$
(black curves), compared with the
leading-order Gaussian distribution
(red dotted curve).
Inset: numerical covariances
$\operatorname{Cov}(x_n^{\mathrm{log}},
x_{n+j}^{\mathrm{log}})$ (circles)
compared with the analytical values
$\kappa_j$ (red line).
(b) Structure functions $S_p(n)$ for
orders $p = 1,\ldots,10$ (circles), with power-law fits
(straight lines).
Inset: measured anomalous corrections
$\zeta_p-p/3$ (circles), compared with the
perturbative result \eqref{eq:scaling_exponents}
(red line).
}
\label{fig2}
\end{figure*}

We compare our analytical results with numerical simulations of the forward cascade
at $a=0.2$ and $\varepsilon=0.1$.
Since the hidden scaling symmetry of the nonlinear
system is expressed in terms of shell multipliers
\cite{mailybaev2021hidden,mailybaev2022shell},
we consider their logarithmic fluctuations
$x_n^{\mathrm{log}}=\varepsilon^{-1}
\log|\lambda^{1/3}u_{n+1}/u_n|$,
which reduce to $x_n$ from
Eq.~\eqref{eq:multiplier_linear}
at leading order in $\varepsilon$.
Figure~\ref{fig2}(a) shows that their PDFs collapse
across shells and closely follow the predicted
Gaussian distribution with zero mean and variance
$\kappa_0$.
The measured covariances also agree with the
analytical solution (inset).
Figure~\ref{fig2}(b) shows power-law scaling of
the structure functions, with anomalous exponents
in excellent agreement with
Eq.~\eqref{eq:scaling_exponents} (inset).
See also SM~\cite[\S V]{SM} for simulations
at $\varepsilon=0.02$ and $0.5$.

\vspace{2mm}
\textit{Cascade transition.}
As the inverse cascade approaches the
transition at $a=0$, velocity fluctuations
become correlated over increasingly large
scale separations.
The asymptotic decay
$c_j-c_\infty\propto r_2^j=e^{-j/\xi_-}$
defines the correlation length
\begin{equation}
    \xi_-=-\frac{1}{\log r_2}
    \sim-\frac{1}{6a},
    \qquad a\nearrow0,
    \label{eq:correlation_length}
\end{equation} 
which diverges
at cascade reversal. 
Simultaneously, for every $j$, the covariance
diverges as (see SM~\cite[\S I\,C]{SM})
\begin{equation}
    c_j-c_\infty
    \sim-\frac{(\rho-1)^2}
    {24\rho(\rho+1)}\,\frac1a,
    \qquad a\nearrow0.
    \label{eq:covariance_divergence}
\end{equation}
In contrast, the multiplier covariances remain
continuous at the cascade transition, with
\begin{equation}
    \kappa_j(0)=
    \begin{cases}
        (\rho+1)/(2\rho), & j=0,\\
        -1/(\rho+1), & j=1,\\
        0, & j\geq2.
    \end{cases}
    \label{eq:kappa_transition}
\end{equation}
Their derivatives, however, exhibit
a finite jump that is independent of
shell separation (see SM~\cite[\S II\,D]{SM}).

The cascade transition has a striking consequence
for anomalous scaling.
The integrated covariance $\mathcal K$
jumps from zero in the inverse cascade
to a finite positive limit on the
forward-cascade side,
\begin{equation}
    \lim_{a\searrow0}\mathcal K
    =\frac{(\rho-1)^2}{2\rho(\rho+1)}>0.
    \label{eq:kappa_sum_transition}
\end{equation}
Hence, the leading anomalous corrections in
Eq.~\eqref{eq:scaling_exponents} change
discontinuously, despite the continuity of
$\kappa_j$ at every fixed shell separation.
This jump originates from the accumulated
contribution of weak long-range negative
correlations in the inverse cascade.

\begin{figure*}
\centering
\includegraphics[width=0.92\textwidth]{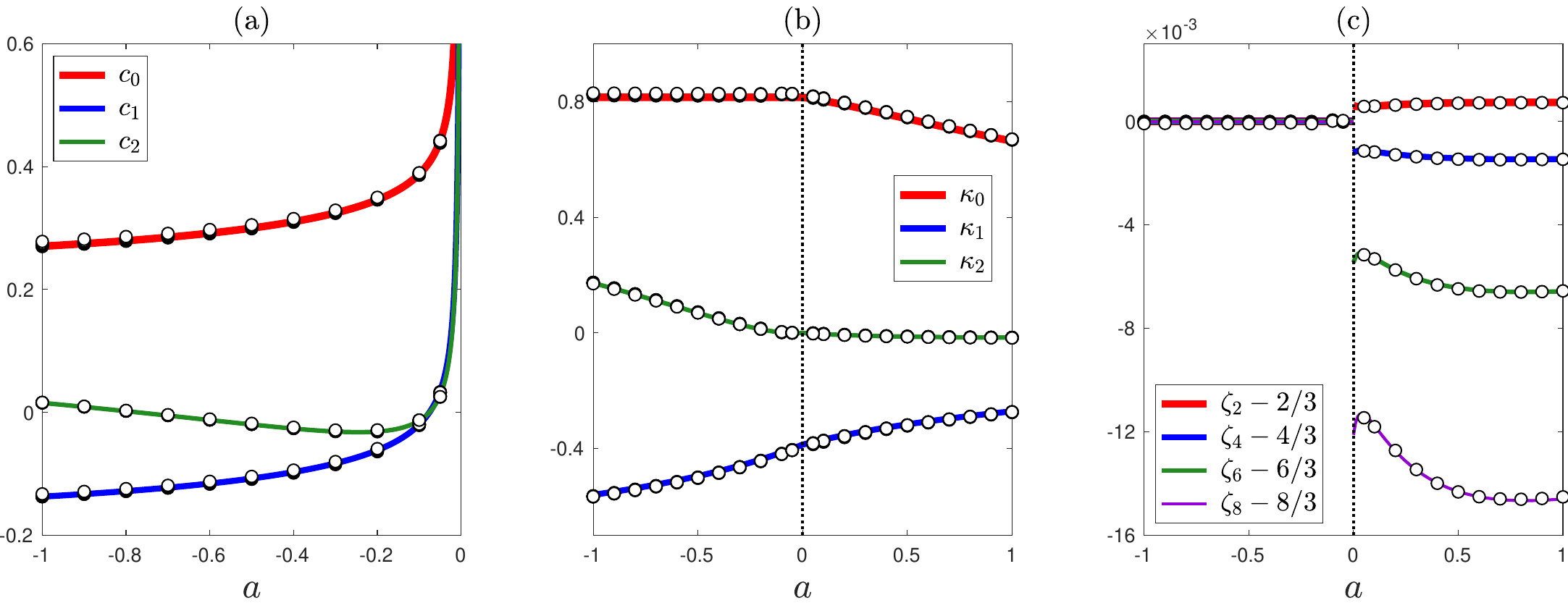}
   \caption{
Statistical transition at cascade reversal
for $\varepsilon=0.1$.
(a) Numerical covariances
$\operatorname{Cov}(v_n,v_{n+j})$
for $n=6,\ldots,14$ and $j=0,1,2$
(circles), compared with the analytical
values $c_j$ (colored lines)
in the inverse cascade.
(b) Numerical multiplier covariances
$\operatorname{Cov}(x_n^{\mathrm{log}},
x_{n+j}^{\mathrm{log}})$
for $n=6,\ldots,14$ and $j=0,1,2$
(circles), compared with the analytical
values $\kappa_j$ (colored lines).
(c) Anomalous corrections to the scaling
exponents, $\zeta_p-p/3$ for $p=2,4,6,8$,
showing the discontinuous onset of intermittency
in the forward cascade: numerical results
(circles) and analytical values
(colored lines).
}
\label{fig3}
\end{figure*}

Figure~\ref{fig3} summarizes the statistical
transition at cascade reversal.
The numerical results show excellent agreement
with the analytical theory.

\vspace{2mm}\textit{Discussion.}
Our results identify a dynamical mechanism
linking cascade direction to statistical
scaling symmetry: the energy flux controls
the growth or decay of fluctuations across
scales, determining whether the K41 symmetry
or only the weaker hidden symmetry of velocity
multipliers \cite{mailybaev2022shell}
is statistically restored. 
Within the perturbative theory, the integrated
multiplier covariance $\mathcal K$ acts as
an order parameter for intermittency.
Its discontinuous onset at cascade reversal,
accompanied by a diverging correlation length,
is reminiscent of mixed-order phase transitions
\cite{bar2014mixed}.
Our analytical approach opens a path toward
a first-principles understanding of transitions
between intermittent and nonintermittent regimes
in other cascade models \cite{benzi2005dynamical}
and ultimately in fluid turbulence
\cite{alexakis2018cascades}. 
\\

\noindent\textit{Acknowledgments.}
I am particularly grateful to Luca Biferale
for stimulating discussions that helped
motivate this study, and to Massimo Cencini
and Alexandros Alexakis for valuable conversations
and encouragement.
This work was supported by CNPq grant
300704/2026-7 and the CAPES MATH-AmSud
project CHA${}^2$MAN.

\vspace{2mm}
\noindent\textit{Data availability.}
The scripts used to generate the figures and numerical results in this work are publicly available at 
\url{https://doi.org/10.5281/zenodo.24559781}.

\vspace{2mm}
\noindent\textit{AI use disclosure.}
ChatGPT (GPT-6) was used to assist
with checking mathematical derivations
and improving the scientific presentation.
The author independently verified the
mathematical results.

\bibliographystyle{unsrt}

\bibliography{biblio}

\clearpage
\onecolumngrid

\setcounter{page}{1}
\setcounter{equation}{0}
\setcounter{figure}{0}
\setcounter{table}{0}
\setcounter{section}{0}

\renewcommand{\theequation}{SM.\arabic{equation}}
\renewcommand{\thefigure}{SM\arabic{figure}}

\begin{center}
{\Large \textbf{Supplemental Material}}

\vspace{0.3cm}

{\large \bf Why is the forward cascade intermittent but the inverse cascade not?}

\vspace{0.3cm}

Alexei A. Mailybaev

\vspace{0.2cm}

IMPA, Rio de Janeiro, Brazil

\end{center}

\vspace{0.5cm}

This Supplemental Material presents the
derivation of stationary covariances,
the perturbative calculation of anomalous
scaling exponents, numerical methods
and additional simulation results.

\section{Stationary covariances of velocity fluctuations $v_n$}
\label{SM1v}

\subsection{Lyapunov equation}  \label{sec_SM_A1}

We consider the infinite-dimensional
Ornstein--Uhlenbeck (OU) system
\begin{equation}
    \rho^{-n}dv_n
    +3\Pi_*\frac{v_{n+1}-v_{n-1}}{2}\,dt
    =(v_{n+1}-2v_n+v_{n-1})\,dt
    +\rho^{-n/2}
    \big[dW_{n-1}-\rho^{-1/2}dW_n\big],
    \qquad n \in \mathbb{Z},
    \label{eqSM:OU_system}
\end{equation}
where $\rho=\lambda^{2/3}>1$ and
$\Pi_*=2a$.
In matrix notation, this system can be written as
\begin{equation}
    dv=\mathcal L v\,dt+\Sigma\,dW,
    \label{eqSM:OU_operator}
\end{equation}
where $v=(v_n)_{n\in\mathbb Z}$ and
$W=(W_n)_{n\in\mathbb Z}$.
The deterministic tridiagonal operator $\mathcal L$
and the bidiagonal noise coefficient matrix $\Sigma$
have elements
\begin{equation}
    \mathcal L_{nm}
    =\rho^n\bigg(
        \delta_{m,n+1}-2\delta_{m,n}
        +\delta_{m,n-1}
        -3\Pi_*
        \frac{\delta_{m,n+1}-\delta_{m,n-1}}{2}
    \bigg),\qquad
    \Sigma_{nm}
    =\rho^{n/2}\big(
        \delta_{m,n-1}
        -\rho^{-1/2}\delta_{m,n}
    \big).
    \label{eqSM:OU_matrices}
\end{equation}

We seek centered stationary Gaussian statistics
characterized by the covariance matrix
$C_{ij}=\langle v_i v_j\rangle_0$,
where $\langle\cdot\rangle_0$ denotes
averaging with respect to the Gaussian measure.
Stationarity requires the covariance matrix
to satisfy the Lyapunov equation
\begin{equation}
    \mathcal LC+C\mathcal L^{\mathsf T}
    +\Sigma\Sigma^{\mathsf T}=0.
    \label{eqSM:OU_Lyapunov}
\end{equation}
As described in the Letter, Gaussian statistics invariant under the
K41 scaling symmetry must have a Toeplitz
covariance $C_{ij}=c_{i-j}$, with $c_{-j}=c_j$.
By this symmetry, it suffices to consider
the components of \eqref{eqSM:OU_Lyapunov}
with $i=0$ and $j\geq0$.
Substituting the matrix elements
\eqref{eqSM:OU_matrices}, we recast the
Lyapunov equation as
\begin{equation}
    3\Pi_*\frac{\rho^j-1}{\rho^j+1}
    \frac{c_{j+1}-c_{j-1}}{2}
    =c_{j+1}-2c_j+c_{j-1}
    +\frac{(1+\rho^{-1})\delta_{j0}
    -\delta_{j1}}{\rho^j+1},
    \qquad j\geq0,
    \label{eqSM:c_recurrence}
\end{equation}
which forms an infinite linear system
for the coefficients $c_j$, $j\geq0$,
with $c_{-1}=c_1$.

For large shell separations $j\to\infty$,
the transport coefficient in
\eqref{eqSM:c_recurrence} approaches
$3\Pi_*$, while the stochastic contribution
vanishes.
The covariance equation therefore reduces to
\begin{equation}
    3\Pi_*\frac{c_{j+1}-c_{j-1}}{2}
    \simeq c_{j+1}-2c_j+c_{j-1},
    \qquad j\to\infty.
    \label{eqSM:c_asymptotic}
\end{equation}
Substituting $c_j=r^j$ into the limiting
equation and using $\Pi_*=2a$ yields
the characteristic equation
\begin{equation}
    3a(r^2-1)=r^2-2r+1,
\end{equation}
with roots
\begin{equation}
    r_1=1,
    \qquad
    r_2=\frac{1+3a}{1-3a}.
    \label{eqSM:stationary_eigenvalues}
\end{equation}

\subsection{Analytic solution}

We now solve the Lyapunov system
\eqref{eqSM:c_recurrence} analytically
for the inverse cascade $a<0$.
Introducing the differences
$d_j=c_j-c_{j-1}$ and using $c_{-1}=c_1$,
the equation at $j=0$ gives
\begin{equation}
    d_1=-\frac{1+\rho^{-1}}{4}.
    \label{eqSM:d1}
\end{equation}
For $j\geq1$, the recurrence takes the form
\begin{equation}
    [\rho^j(1-3a)+(1+3a)]d_{j+1}
    -[\rho^j(1+3a)+(1-3a)]d_j
    =\delta_{j1}.
    \label{eq:d_recurrence_SM}
\end{equation}
The equation at $j=1$ determines
\begin{equation}
    d_2=d_1\,
    \frac{(\rho-1)^2+3a(\rho^2-1)}
         {(\rho+1)^2-3a(\rho^2-1)}.
    \label{eqSM:d2}
\end{equation}
For $j\geq2$, the homogeneous recurrence
yields
\begin{equation}
    d_j=d_2\prod_{m=2}^{j-1}
    \frac{\rho^m(1+3a)+(1-3a)}
         {\rho^m(1-3a)+(1+3a)},
    \qquad j\geq3.
    \label{eqSM:covariance_d_solution}
\end{equation}
The factors in this product approach the
flux-preserving root
$r_2=(1+3a)/(1-3a)$ as $m\to\infty$.
Since $|r_2|<1$ for $a<0$,
the differences $d_j$ decay and are
absolutely summable.
The bounded covariance solutions are therefore
\begin{equation}
    c_j=c_\infty-\sum_{m=j+1}^{\infty}d_m,
    \qquad j\geq0,
    \label{eqSM:covariance_solution}
\end{equation}
where $c_\infty=\lim_{j\to\infty}c_j$
is a free constant corresponding to the
variance of a fluctuation common to all shells.
Positive definiteness requires $c_\infty\geq0$.

Finally, let us prove that K41-invariant
stationary Gaussian statistics are impossible
for $a\geq0$ by showing that the Lyapunov
equation admits no bounded solution. 
Indeed, unless the recurrence
\eqref{eq:d_recurrence_SM} is inconsistent
due to a vanishing coefficient of $d_{j+1}$,
its solutions with \eqref{eqSM:d1}
satisfy $d_2\neq0$ and
$|d_{j+1}/d_j|\geq1$ for $j\geq2$,
with $d_{j+1}/d_j\to r_2$ as $j\to\infty$.
At $a=0$, we have $r_2=1$ and $d_j=d_2<0$
for $j\geq2$, so $c_j$ decreases linearly with $j$.
For $a>0$ and $a\neq1/3$, we have $|r_2|>1$,
so the differences $d_j$ grow exponentially
in absolute value, making $c_j$ unbounded.
At $a=1/3$, we have $r_2=\infty$,
and the ratio $d_{j+1}/d_j=\rho^j$
yields superexponential growth.

\subsection{Asymptotic properties}

We examine the covariance asymptotics
for $a<0$ sufficiently close to the transition,
where $0<r_2<1$.
Extracting the limiting factor $r_2$ from
each term in the product
\eqref{eqSM:covariance_d_solution}, we obtain
\begin{equation}
    d_j=d_2 r_2^{j-2}Q_j,
    \qquad
    Q_j=\prod_{m=2}^{j-1}
    \frac{1+\rho^{-m}/r_2}
         {1+r_2\rho^{-m}}.
    \label{eqSM:d_asymptotic_product}
\end{equation}
As $r_2=1+O(|a|)$, the factors in this
product differ from unity by
$O(|a|\rho^{-m})$.
Hence, the product converges to a finite
positive limit $Q_\infty$, with
\begin{equation}
    Q_\infty=1+O(|a|),
    \qquad
    Q_j-Q_\infty=O(|a|\rho^{-j}).
    \label{eq_Q_limit}
\end{equation}
Substituting \eqref{eqSM:d_asymptotic_product}
into \eqref{eqSM:covariance_solution} and
using \eqref{eq_Q_limit}, we obtain
\begin{equation}
    c_j-c_\infty
    =-d_2\sum_{m=j+1}^{\infty}r_2^{m-2}Q_m
    =-\frac{d_2Q_\infty r_2^{j-1}}{1-r_2}
    +O(|a|r_2^{j-1}\rho^{-j}),
    \qquad j\geq1.
    \label{eqSM:c_large_j}
\end{equation}
From the explicit expressions for $d_2$
and $r_2$, we have
\begin{equation}
    d_2 \to
    -\frac{(\rho-1)^2}{4\rho(\rho+1)},
    \qquad
    1-r_2\sim-6a,
    \qquad a\nearrow0.
    \label{eq_limits}
\end{equation}
Since $Q_\infty\to1$ and $r_2^{j-1}\to1$
for every fixed $j$, Eq.~\eqref{eqSM:c_large_j}
yields
\begin{equation}
    c_j-c_\infty
    \sim-\frac{(\rho-1)^2}
    {24\rho(\rho+1)}\,\frac1a,
    \qquad a\nearrow0.
    \label{eqSM:c_divergence}
\end{equation}
This divergence is the same for every fixed
shell separation, including $j=0$ because
$c_0=c_1-d_1$ and $d_1$ is finite.

\section{Stationary covariances of multiplier fluctuations $x_n$}
\label{SM1x}

\subsection{Lyapunov equation} 

The infinite-dimensional Ornstein--Uhlenbeck
(OU) system for the multiplier fluctuations
$x_n=v_{n+1}-v_n$ is obtained by subtracting
Eq.~\eqref{eqSM:OU_system} at indices
$n+1$ and $n$.
It takes the conservative form
\begin{equation}
    dx_n=(\Phi_{n-1}-\Phi_n)\,dt
    -\rho^{n/2}
    \big[dW_{n-1}
    -(\rho^{1/2}+\rho^{-1/2})dW_n
    +dW_{n+1}\big],
    \label{eqSM:OU_x}
\end{equation}
where $\Phi_n$ is the flux of multiplier
fluctuations,
\begin{equation}
    \Phi_n=\rho^{n+1}\left[
        \frac{3\Pi_*}{2}(x_n+x_{n+1})
        -(x_{n+1}-x_n)
    \right].
    \label{eqSM:multiplier_flux}
\end{equation}
We seek stationary Gaussian statistics
with Toeplitz covariance
$K_{ij}=\langle x_i x_j\rangle_0
=\kappa_{i-j}$, where $\kappa_{-j}=\kappa_j$.
Following the procedure of
Sec.~\ref{sec_SM_A1}, we obtain the
Lyapunov equation
\begin{equation}
    \begin{aligned}
    &\big[\rho^j(1+3a)+\rho(1-3a)\big]
        \kappa_{j-1}
     -(1+\rho^j)
        \big[\rho(1+3a)+(1-3a)\big]\kappa_j
    +\big[(1+3a)+\rho^{j+1}(1-3a)\big]
        \kappa_{j+1}\\
    &=- (\rho+\rho^{-1}+4)\delta_{j0}
        +2(\rho+1)\delta_{j1}
        -\rho\delta_{j2},
        \qquad j\geq0.
    \end{aligned}
    \label{eq:kappa_recurrence}
\end{equation}
Dividing \eqref{eq:kappa_recurrence}
by $\rho^j$ and retaining the leading
terms as $j\to\infty$, we obtain
\begin{equation}
    (1+3a)\kappa_{j-1}
    -\big[\rho(1+3a)+(1-3a)\big]\kappa_j
    +\rho(1-3a)\kappa_{j+1}
    \simeq0,
    \qquad j\to\infty.
    \label{eqSM:kappa_asymptotic}
\end{equation}
Substituting $\kappa_j=r^j$ into the limiting
equation yields the characteristic equation
\begin{equation}
    (1+3a)
    -\big[\rho(1+3a)+(1-3a)\big]r
    +\rho(1-3a)r^2=0,
\end{equation}
with roots
\begin{equation}
    r_1'=\rho^{-1},
    \qquad
    r_2'=\frac{1+3a}{1-3a}=r_2.
\end{equation}
As compared with the roots
\eqref{eqSM:stationary_eigenvalues}
for velocity covariances,
the marginal root $r_1=1$ is replaced
by the decaying root $r_1'=\rho^{-1}<1$,
while the flux-preserving root $r_2$
remains unchanged.

\subsection{Analytic solution for the forward cascade}

We now solve the Lyapunov system
\eqref{eq:kappa_recurrence} analytically
for the forward cascade $a>0$.
Boundedness selects the decaying spatial
mode $r_1'=\rho^{-1}<1$, so that
$\kappa_j\propto\rho^{-j}$ as $j\to\infty$.

We first seek a solution $g_0,g_1,g_2,\ldots$
of the homogeneous recurrence
\begin{equation}
    \big[\rho^j(1+3a)+\rho(1-3a)\big]g_{j-1}
    -(1+\rho^j)\big[\rho(1+3a)+(1-3a)\big]g_j
    +\big[(1+3a)+\rho^{j+1}(1-3a)\big]g_{j+1}=0,
     \quad j\geq1,
    \label{eq:g_recurrence}
\end{equation}
normalized by $g_j\sim\rho^{-j}$ as $j\to\infty$.
We use a power-series expansion in $\rho^{-j}$,
\begin{equation}
    g_j=\rho^{-j}
    \sum_{m=0}^{\infty}h_m\rho^{-mj},
    \qquad h_0=1.
    \label{eq:g_series}
\end{equation}
Substituting this expansion into
\eqref{eq:g_recurrence} and matching
powers of $\rho^{-j}$ yields
\begin{equation}
    h_m=
    \frac{\rho^{m+1}-1}{\rho^m-1}
    \frac{(1+3a)-(1-3a)\rho^m}
         {(1+3a)\rho^{m+1}-(1-3a)}
    h_{m-1},
    \qquad m\geq1.
    \label{eq_rec_h}
\end{equation}
Iterating this relation, we obtain
\begin{equation}
    h_m=
    \frac{\rho^{m+1}-1}{\rho-1}
    \prod_{\ell=1}^{m}
    \frac{
        (1+3a)-(1-3a)\rho^\ell
    }{
        (1+3a)\rho^{\ell+1}-(1-3a)
    },
    \qquad m\geq1.
    \label{eq:g_coefficients}
\end{equation}
The series converges for all $j\geq0$,
since the multiplicative factor in the
recurrence for $h_m$ approaches
$-(1-3a)/(1+3a)$, whose absolute value
is less than unity for $a>0$.

Since the right-hand side of
\eqref{eq:kappa_recurrence} vanishes for
$j\geq3$, the bounded covariance takes the form
\begin{equation}
    \kappa_j=\eta g_j,
    \qquad j\geq2,
    \label{eq:kappa_tail}
\end{equation}
with an undetermined constant $\eta$.
The remaining unknowns $\eta$, $\kappa_1$,
and $\kappa_0$ are fixed by
\eqref{eq:kappa_recurrence} at $j=2,1,0$.
The equations at $j=2$ and $j=1$ yield
\begin{equation}
    \kappa_1
    =\eta g_1
    -\frac{1}
    {\rho(1+3a)+(1-3a)},\qquad
    \kappa_0
    =\eta g_0
    +\frac{\rho+1}{2\rho}.
    \label{eq:kappa_boundary}
\end{equation}
Finally, the equation at $j=0$
determines
\begin{equation}
    \eta=
    \frac{
        3a(\rho-1)^2
        \big[(\rho-1)+3a(\rho+1)\big]
    }{
        2\rho\big[\rho(1+3a)+(1-3a)\big]
        \left[
        \big[(1+3a)+\rho(1-3a)\big]g_1
        -\big[\rho(1+3a)+(1-3a)\big]g_0
        \right]
    }.
    \label{eq:kappa_eta}
\end{equation}
Thus, Eqs.~\eqref{eq:g_series}--\eqref{eq:kappa_eta}
determine the unique Toeplitz
covariance solution for $a>0$.

The integrated multiplier covariance
\begin{equation}
    \mathcal K
    =\sum_{j\in\mathbb Z}\kappa_j
    =\kappa_0+2\sum_{j=1}^{\infty}\kappa_j
    \label{eq:K_sum}
\end{equation}
can be evaluated directly from the Lyapunov
recurrence without summing the explicit
covariance solution.
Multiplying \eqref{eq:kappa_recurrence}
by $\rho^{-j}$, summing over $j\geq1$,
and using the equation at $j=0$ to
eliminate $\kappa_1$, we obtain
\begin{equation}
    \mathcal K=
    \frac{1+\rho^{-1}-2\kappa_0}{6a}
    =-\frac{\eta g_0}{3a},
    \qquad a>0,
    \label{eq:K_explicit}
\end{equation}
where the last equality follows from
\eqref{eq:kappa_boundary}.

\subsection{Analytic solution for the inverse cascade}

For the inverse cascade $a<0$, both
characteristic roots $r_1'=\rho^{-1}$
and $r_2'=r_2$ have modulus less than
unity. Hence, boundedness alone does not
uniquely determine the multiplier covariance.
We instead select the solution compatible
with K41-invariant velocity statistics,
whose covariance $c_j$ was obtained in
the previous section up to an additive
constant $c_\infty$.
Since $x_n=v_{n+1}-v_n$, the multiplier
covariances satisfy
\begin{equation}
    \kappa_j=2c_j-c_{j-1}-c_{j+1},
    \qquad c_{-j}=c_j.
    \label{eq:kappa_from_c}
\end{equation}
The dependence on $c_\infty$ cancels.
In terms of the differences
$d_j=c_j-c_{j-1}$, we obtain
\begin{equation}
    \kappa_0=-2d_1,
    \qquad
    \kappa_j=d_j-d_{j+1},
    \quad j\geq1.
    \label{eqSM:kappa_from_d}
\end{equation}
Since $d_j\to0$, the integrated covariance
vanishes by telescopic summation,
\begin{equation}
    \mathcal K
    =-2\lim_{j\to\infty}d_j=0,
    \qquad a<0.
    \label{eq:K_inverse}
\end{equation}
Thus, K41 compatibility selects the
multiplier covariance with $\mathcal K=0$,
in contrast to $\mathcal K>0$ in the
forward cascade.

\subsection{Asymptotic properties}

We examine the covariance limits at the
cascade transition $a=0$.
As $a\searrow0$, the multiplicative factor
in the recurrence \eqref{eq_rec_h} approaches $-1$. Since $h_0 = 1$, this 
yields $h_m\to(-1)^m$ for every fixed $m$.
Furthermore, for sufficiently small $a>0$,
this factor lies between $-1$ and $0$,
so that $|h_m|\leq1$.
For $j\geq1$, we can therefore take the limit
term by term in \eqref{eq:g_series},
obtaining
\begin{equation}
    \lim_{a\searrow0}g_j
    =\rho^{-j}\sum_{m=0}^{\infty}(-1)^m\rho^{-mj}
    =\frac{1}{1+\rho^j}.
    \label{eq:g_transition}
\end{equation}
The same expression holds for $j=0$,
as follows from the homogeneous recurrence
\eqref{eq:g_recurrence} at $j=1$. From Eq.~\eqref{eq:kappa_eta}, we have
$\eta\to0$. Therefore, Eqs.~\eqref{eq:kappa_tail}
and \eqref{eq:kappa_boundary} yield
\begin{equation}
    \lim_{a\searrow0}\kappa_j=
    \begin{cases}
        (\rho+1)/(2\rho), & j=0,\\
        -1/(\rho+1), & j=1,\\
        0, & j\geq2.
    \end{cases}
    \label{eq:kappa_forward_limit}
\end{equation}
The uniform exponential decay of the
forward-cascade covariance tails allows
interchanging the limit \eqref{eq:kappa_forward_limit} with the sum in Eq.~\eqref{eq:K_sum},
giving
\begin{equation}
    \mathcal K_*=
    \lim_{a\searrow0}\mathcal K
    =\frac{(\rho-1)^2}{2\rho(\rho+1)}.
    \label{eq:K_transition}
\end{equation}
Using \eqref{eq:K_explicit} and
$\mathcal K\to\mathcal K_*$ as $a\searrow0$,
we obtain
\begin{equation}
    \left.\frac{d\kappa_0}{da}\right|_{0^+}
    =-3\mathcal K_*.
    \label{eq:kappa_forward_derivative}
\end{equation}

For the inverse cascade,
Eqs.~\eqref{eqSM:d1}, \eqref{eqSM:d2} and
\eqref{eqSM:covariance_d_solution} give
$d_1=-(\rho+1)/(4\rho)$ and
$d_j\to-\mathcal K_*/2$ for every fixed
$j\geq2$ as $a\nearrow0$.
Using \eqref{eqSM:kappa_from_d},
we recover the same limits
\eqref{eq:kappa_forward_limit},
establishing continuity of every $\kappa_j$.
Moreover, $\kappa_0=-2d_1$ is independent
of $a$ for $a<0$, so that
$\kappa_0'(0^-)=0$.
Together with \eqref{eq:kappa_forward_derivative},
this establishes the one-sided derivatives
of $\kappa_0$.
The recurrence \eqref{eq:kappa_recurrence}
then implies the existence of one-sided
derivatives for every fixed $j$.

To determine the jump in the derivatives,
we define
\begin{equation}
    \Delta_j=
    \left.\frac{d\kappa_j}{da}\right|_{0^+}
    -\left.\frac{d\kappa_j}{da}\right|_{0^-}.
\end{equation}
Since both covariance solutions have the same
limit at $a=0$, differentiating
\eqref{eq:kappa_recurrence} on both sides
and subtracting yields
\begin{equation}
    (\rho^j+\rho)\Delta_{j-1}
    -(1+\rho^j)(\rho+1)\Delta_j
    +(1+\rho^{j+1})\Delta_{j+1}=0,
    \qquad j\geq1.
    \label{eq:kappa_jump_recurrence}
\end{equation}
The equation at $j=0$ gives
$\Delta_1=\Delta_0$, and the recurrence
then implies $\Delta_j=\Delta_0$
for all $j\geq0$.
Hence,
\begin{equation}
    \Delta_j=-3\mathcal K_*
    =-\frac{3(\rho-1)^2}{2\rho(\rho+1)},
    \qquad j\geq0.
    \label{eq:kappa_jump_SM}
\end{equation}

In summary, all multiplier covariances remain
continuous at cascade reversal, while
their derivatives exhibit the same finite
jump at every fixed shell separation.
In contrast, their integrated sum changes
discontinuously from $\mathcal K=0$ in
the inverse cascade to $\mathcal K_*>0$
on the forward-cascade side.

\section{Anomalous exponents in the forward cascade}
\label{sec:intermittent_forward}

Here we present a formal perturbative
derivation of the leading anomalous scaling
exponents from the stationary multiplier
covariances and the constant energy flux.
For convenience, we place the infrared cutoff
at $n=0$ and fix $u_0=1$,
neglecting cutoff effects on the inertial-range
multiplier statistics. We also perform the perturbative expansion
in the positive-amplitude sector $u_n>0$,
neglecting sign changes whose probabilities
are exponentially small in the leading
Gaussian approximation at fixed $n$.

We extend the perturbative expansion
of the multipliers to second order,
\begin{equation}
    \frac{u_{n+1}}{u_n}
    =\lambda^{-1/3}
    \left[
        1+\varepsilon x_n+\varepsilon^2 y_n
        +O(\varepsilon^3)
    \right],
    \label{eq:multiplier_linear_B}
\end{equation}
where $x_n$ is the leading-order Gaussian
fluctuation and $y_n$ is the second-order
correction to the stochastic multiplier.
The telescoping product of multipliers gives
\begin{equation}
    u_n
    =\prod_{j=0}^{n-1}\frac{u_{j+1}}{u_j}
    =\lambda^{-n/3}
    \bigg[
        1+\varepsilon\sum_{j=0}^{n-1}x_j
        +\varepsilon^2\left(
        \sum_{j=0}^{n-1}y_j
        +\sum_{0\leq i<j<n}x_ix_j
        \right)
        +O(\varepsilon^3)
    \bigg].
    \label{eq:telescopic_expansion}
\end{equation}
Raising this expression to the power $p$
and averaging, we obtain
\begin{equation}
    S_p(n)=\lambda^{-np/3}
    \bigg[1+\varepsilon^2\bigg(
        p\sum_{j=0}^{n-1}\langle y_j\rangle
        +\frac{p(p-1)}{2}n\kappa_0
        +p^2\sum_{r=1}^{n-1}(n-r)\kappa_r
    \bigg)
    +O(\varepsilon^4)\bigg].
    \label{eq:telescopic_moments}
\end{equation}
Here we used the centered Gaussian statistics
of the leading-order fluctuations $x_n$,
with Toeplitz covariance
$\langle x_i x_j\rangle_0=\kappa_{i-j}$.
The averages $\langle y_j\rangle$ represent
the mean second-order corrections.
Odd powers of $\varepsilon$ vanish in the
averaged perturbative expansion because
the stochastic dynamics is invariant in law
under $\varepsilon\mapsto-\varepsilon$.

Since the multiplier covariances decay
exponentially, for large $n$ we have
\begin{equation}
    \sum_{r=1}^{n-1}(n-r)\kappa_r
    =n\sum_{r=1}^{\infty}\kappa_r+o(n)
    =\frac{n}{2}(\mathcal K-\kappa_0)+o(n).
    \label{eq:telescopic_covariance_sum}
\end{equation}
We define the asymptotic mean of the
second-order multiplier corrections as
\begin{equation}
    \bar y=\lim_{n\to\infty}
    \frac{1}{n}\sum_{j=0}^{n-1}
    \langle y_j\rangle.
    \label{eq:telescopic_y_mean}
\end{equation}
Substituting into \eqref{eq:telescopic_moments}
yields
\begin{equation}
    \begin{aligned}
    S_p(n)=\lambda^{-np/3}
    \bigg[1+\varepsilon^2 n
    \left(
        p\bar y
        +\frac{p^2\mathcal K-p\kappa_0}{2}
    \right)
    +\varepsilon^2 o(n)+O(\varepsilon^4)
    \bigg].
    \end{aligned}
    \label{eq:telescopic_moments_asymptotic}
\end{equation}
Here the expansion in $\varepsilon$ is
understood at fixed $n$, while $o(n)$
refers to the limit $n\to\infty$.

We assume a regular perturbative expansion
of the inertial-range scaling exponents,
\begin{equation}
    \zeta_p=\frac{p}{3}
    +\varepsilon^2\zeta_p^{(2)}
    +O(\varepsilon^4).
    \label{eq:zeta_expansion}
\end{equation}
The inertial-range power law $S_p(n)\propto\lambda^{-n\zeta_p}$
then yields, upon expansion in $\varepsilon$,
\begin{equation}
    S_p(n)=C_p\lambda^{-np/3}
    \left[
        1-\varepsilon^2n\zeta_p^{(2)}
        \log\lambda+O(\varepsilon^4)
    \right].
    \label{eq:scaling_taylor}
\end{equation}
Comparing the leading terms proportional to
$n\varepsilon^2$ in
\eqref{eq:telescopic_moments_asymptotic}
and \eqref{eq:scaling_taylor}, we obtain
\begin{equation}
    \zeta_p^{(2)}
    =-\frac{p}{2\log\lambda}
    \left(p\mathcal K+2\bar y-\kappa_0\right).
    \label{eq:telescopic_exponents}
\end{equation}

The quantity $\bar y$ can be computed directly
using a higher-order perturbative expansion
of the multiplier distribution
\cite{mailybaev2026perturbative,mailybaev2026first}. 
Here we avoid this calculation by exploiting
the constant-flux constraint $\zeta_3=1$
for the inertial-range statistics.
Setting $\zeta_3^{(2)}=0$
in \eqref{eq:telescopic_exponents} yields
$\bar y= (\kappa_0-3\mathcal K)/2$.
Substituting this result back into
\eqref{eq:telescopic_exponents} and \eqref{eq:zeta_expansion}, we obtain
\begin{equation}
    \zeta_p=\frac{p}{3}
    -\frac{\varepsilon^2p(p-3)}
          {2\log\lambda}\mathcal K
    +O(\varepsilon^4).
    \label{eq:HS_anomalous_exponents}
\end{equation}

Thus, the leading anomalous exponents are
determined entirely by the Gaussian multiplier
covariances and the constant-flux constraint.
More generally, the hidden-symmetry formalism
\cite{mailybaev2022shell,mailybaev2026perturbative,
mailybaev2026first}
and its associated Perron--Frobenius theory
provide a systematic perturbative framework
for deriving inertial-range power laws and
their anomalous exponents from first principles
in this class of models.

\section{Numerical simulations}
\label{app:numerics}

\subsection{Conversion to It\^o form}
\label{app:Ito}

We first derive the It\^o form of the ideal
stochastic mixed model.
The Stratonovich equations can be written as
\begin{equation}
    du_n=\mathcal N_n(u)\,dt
    +\varepsilon\sum_j G_{nj}(u)
    \circ dW_j,
    \label{eq:app_Strato}
\end{equation}
where the deterministic drift is
\begin{equation}
    \mathcal N_n(u)
    =k_{n-1}u_{n-1}\left[
        (1+a)\lambda^{1/3}u_{n-1}
        -(1-a)\lambda^{2/3}u_n
    \right]
    -k_nu_{n+1}\left[
        (1+a)\lambda^{1/3}u_n
        -(1-a)\lambda^{2/3}u_{n+1}
    \right],
    \label{eq:app_Strato_N}
\end{equation}
and the noise coefficients are
\begin{equation}
    G_{nj}(u)=
    u_{n-1}\sqrt{k_{n-1}|u_{n-1}|}
        \,\delta_{j,n-1}
    -u_{n+1}\sqrt{k_n|u_n|}
        \,\delta_{j,n}.
    \label{eq:app_G}
\end{equation}
The Stratonovich-to-It\^o correction is
\begin{equation}
    \begin{aligned}
    \mathcal I_n(u)
    &=\frac12\sum_{j,l}
    G_{lj}(u)\frac{\partial G_{nj}(u)}
    {\partial u_l}\\
    &=\frac14\big[
        -3k_{n-1}|u_{n-1}|u_n
        +k_n\operatorname{sgn}(u_n)u_{n+1}^2
        -2k_nu_n|u_n|
    \big].
    \end{aligned}
    \label{eq:app_Ito_correction_explicit}
\end{equation}
Hence, the It\^o equations take the form
\begin{equation}
    du_n=
    \big[\mathcal N_n(u)
        +\varepsilon^2\mathcal I_n(u)\big]dt
    +\varepsilon\big[
        u_{n-1}\sqrt{k_{n-1}|u_{n-1}|}
            \,dW_{n-1}
        -u_{n+1}\sqrt{k_n|u_n|}
            \,dW_n
    \big].
    \label{eq:app_Ito_model}
\end{equation}
The It\^o correction is derived for nonzero
shell amplitudes. At zero, we adopt the
convention $\operatorname{sgn}(0)=0$.

\subsection{Forcing and dissipation}

We simulate the stochastic mixed model on
a finite set of shells $n=1,\ldots,N$
with $\lambda=2$.
We consider two boundary configurations:
\begin{align}
    \mathrm{(F)}:\quad&
    u_0=1,\quad u_{N+1}=0,
    &&\text{large-scale forcing},
    \label{eq:num_boundary_a}\\
    \mathrm{(I)}:\quad&
    u_0=0,\quad
    u_{N+1}=\lambda^{-(N+1)/3},
    &&\text{small-scale forcing}.
    \label{eq:num_boundary_b}
\end{align}
These configurations sustain forward
and inverse energy cascades for $a>0$
and $a<0$, respectively.
Both configurations are also considered
over the full range of $a$ to examine
the cascade transition.

The Stratonovich system at shells $n=1,\ldots,N$ reads
\begin{equation}
    du_n=u_{n-1}\circ dF_{n-1}
         -u_{n+1}\circ dF_n
         -\delta_{n,n_d}Dk_nu_n|u_n|\,dt,
    \label{eq:num_model}
\end{equation}
where $dF_n$ are the stochastic transfer
rates of the ideal model. Here the dissipative term is added at a single shell $n_d$ with
\begin{equation}
    D=\max(2|a|,0.1), \qquad
        n_d=
    \begin{cases}
        N, & \mathrm{(F)},\\
        1, & \mathrm{(I)}.
    \end{cases}
    \label{eq:num_dissipation}
\end{equation}
The choice $D=2|a|$ matches the magnitude
of the K41 energy flux $\Pi_*=2a$,
minimizing cutoff effects.
Indeed, for $\varepsilon=0$, the corresponding
cascade configurations then admit the exact
stationary K41 solution $u_n^*=\lambda^{-n/3}$.
The lower bound $D\geq0.1$ maintains
nonzero dissipation near cascade reversal,
where the K41 flux vanishes.

For numerical integration, we use
the It\^o form of \eqref{eq:num_model},
\begin{equation}
    du_n=
    \big[
        \mathcal N_n(u)
        -\delta_{n,n_d}Dk_nu_n|u_n|
        +\varepsilon^2\mathcal I_n^{\mathrm{cut}}(u)
    \big]dt
    +\varepsilon\sum_jG_{nj}(u)\,dW_j,
    \label{eq:num_Ito}
\end{equation}
where $\mathcal N_n$ and $G_{nj}$
are given in \eqref{eq:app_Strato_N}
and \eqref{eq:app_G}, with fixed
boundary amplitudes $u_0$ and $u_{N+1}$.
Since these amplitudes are not dynamical,
derivatives with respect to them are
excluded from the Stratonovich-to-It\^o
correction.
Consequently,
\begin{equation}
    \mathcal I_n^{\mathrm{cut}}(u)
    =\mathcal I_n(u)
    +\frac34\delta_{n1}k_0|u_0|u_1
    +\frac12\delta_{nN}k_Nu_N|u_N|,
    \label{eq:num_Ito_correction}
\end{equation}
where $\mathcal I_n(u)$ is defined
in \eqref{eq:app_Ito_correction_explicit}.

\subsection{Numerical method and measurements}

We integrate the It\^o system using the
Euler--Maruyama method with independent
Gaussian increments
$\Delta W_n\sim\mathcal N(0,\Delta t)$.
The time step is
\begin{equation}
    \Delta t=0.01\tau_K,
    \qquad \tau_K=k_N^{-2/3},
    \label{eq:num_dt}
\end{equation}
where $\tau_K$ is the K41 turnover time
at the smallest resolved scale.
The initial conditions are
\begin{equation}
    u_n(0)=k_n^{-1/3}(1+0.2\xi_n),
    \qquad \xi_n\sim\mathcal N(0,1),
    \label{eq:num_initial}
\end{equation}
with independent standard Gaussian variables
$\xi_n$.

We measure the normalized velocity
fluctuations as $v_n = (k_n^{1/3}u_n-1)/\varepsilon$.
Our perturbation theory also uses
the differences $x_n=v_{n+1}-v_n$ for the intermittent state.
However, the hidden scaling symmetry
\cite{mailybaev2022shell,mailybaev2026perturbative,
mailybaev2026first}
of the full nonlinear system is expressed
in terms of shell multipliers $u_{n+1}/u_n$.
We therefore introduce the logarithmic variables
\begin{equation}
    x_n^{\mathrm{log}}
    =\frac{1}{\varepsilon}
    \log\left|\lambda^{1/3}\frac{u_{n+1}}{u_n}\right|
    =x_n+O(\varepsilon),
    \label{eq:xn_alternatives}
\end{equation}
which are used for all numerical measurements
of multiplier statistics.
The logarithmic definition also avoids
potential divergences of multiplier variances
due to near-zero denominators.

\section{Simulation results}

To compute stationary statistics, we discard
an initial transient of duration
$T_{\mathrm{tr}}=50$ and collect data over
a subsequent interval $T\approx2000$
for most simulations.
The statistical uncertainties of averaged
quantities are typically smaller than the
plotting symbols, so error bars are omitted.
The main source of uncertainty is expected
to be systematic finite-size effects due
to the forcing and dissipation cutoffs.
Although difficult to quantify, these effects
should be small when the inter-shell correlation length
is much shorter than the available inertial
range.
Most simulations use $N=22$ shells.
However, the correlation length diverges
as the inverse cascade approaches the transition.
We therefore use $N=30$ shells for
$a=-0.1$, $-0.05$, and $-0.03$ to reduce
finite-size effects, with shorter sampling
intervals due to the increased computational cost.

\begin{figure}[p]
\centering
\includegraphics[width=0.8\textwidth]{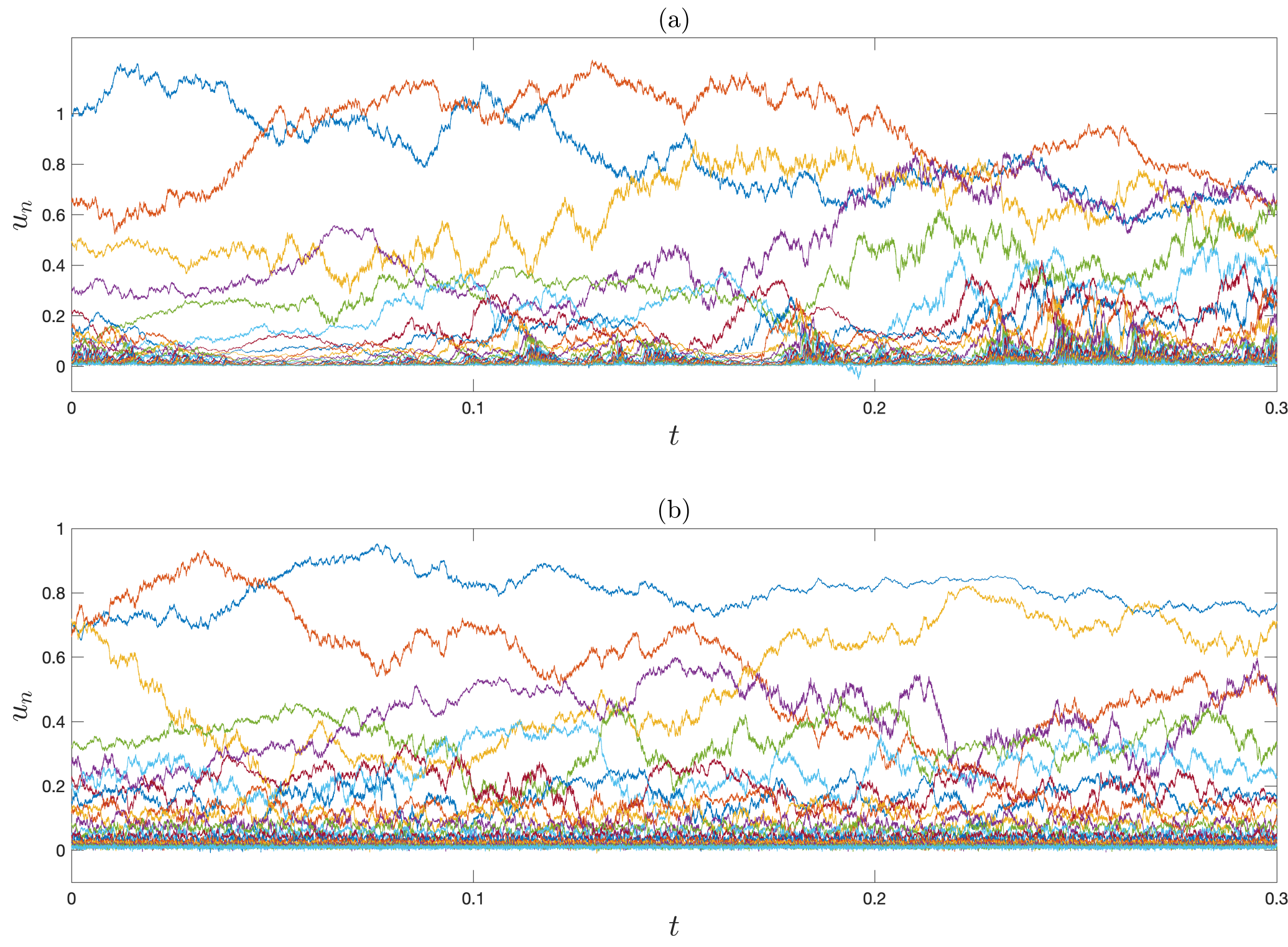}
\caption{
Representative time evolution for
(a) the intermittent forward cascade at $a=1$
and (b) the nonintermittent K41 inverse
cascade at $a=-1$.
Both simulations use $\varepsilon=0.5$
and $N=22$.
}
\label{figSM1}
\end{figure}

Figure~\ref{figSM1} illustrates the contrasting
dynamics of the two cascade regimes.
The forward cascade exhibits pronounced
intermittent bursts, whereas fluctuations
in the inverse cascade are more uniform.

\begin{figure}[p]
\centering
\includegraphics[width=0.99\textwidth]{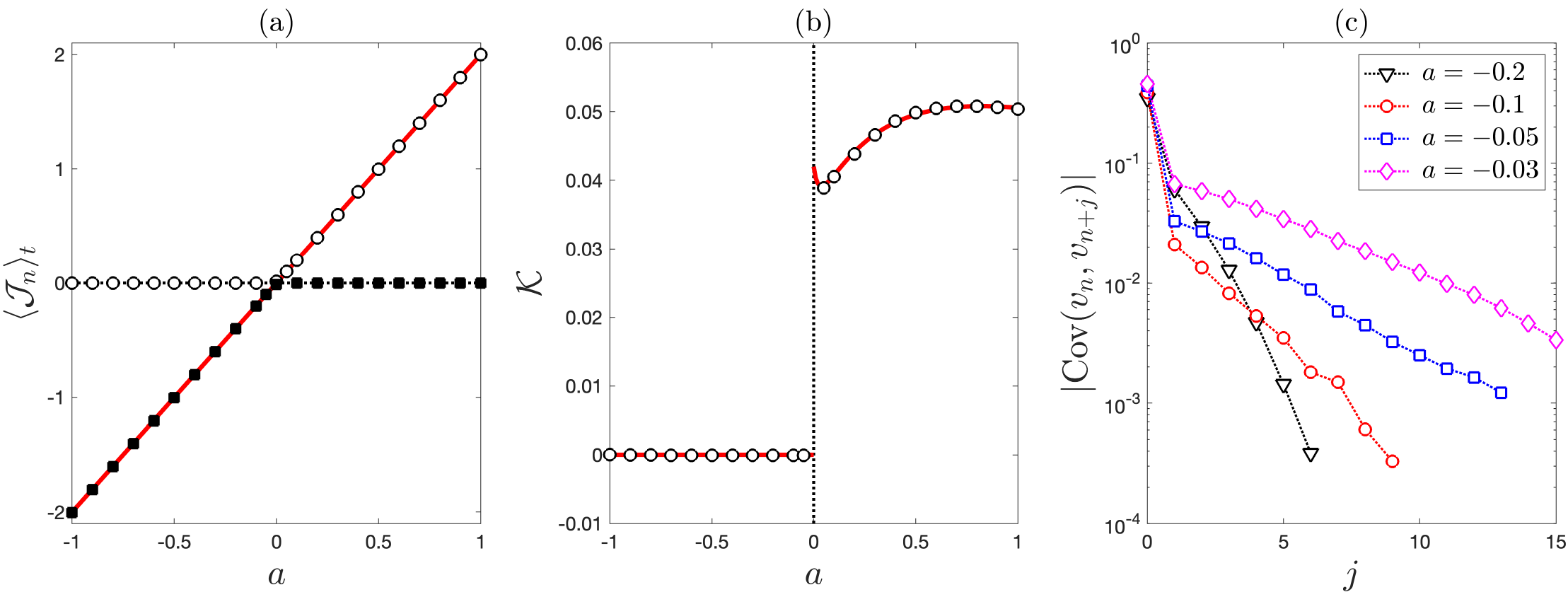}
\caption{
(a) Mean energy flux at shell $n=12$
versus $a$ for large-scale forcing
(open circles) and small-scale forcing
(filled squares), compared with the
K41 value $\Pi_*=2a$ (red line).
(b) Integrated multiplier covariance
$\mathcal K$ versus $a$: analytical
values (red curve) and numerical
estimates (circles) at $n=12$.
(c) Absolute values of velocity covariances
$\operatorname{Cov}(v_n,v_{n+j})$
for $n=12$ and different shell separations $j$
in the inverse cascade.
All panels correspond to
$\varepsilon=0.1$, with $N=22$ except
for $a=-0.1$, $-0.05$, and $-0.03$,
where $N=30$.
}
\label{figSM2}
\end{figure}

The main Letter presents numerical results
at $\varepsilon=0.1$ for different values
of $a$.
Here we provide additional results across
the cascade transition.
Figure~\ref{figSM2}(a) shows the mean
energy flux for both boundary configurations
\eqref{eq:num_boundary_a} and
\eqref{eq:num_boundary_b}.
The flux is close to its K41 value
$\Pi_*=2a$ when forcing is applied at
the appropriate end of the cascade
and is negligible otherwise.
Figure~\ref{figSM2}(b) shows the integrated
multiplier covariance $\mathcal K$.
The numerical estimates are obtained using the measured covariances $\kappa_j\approx
    \operatorname{Cov}(x_n^{\mathrm{log}},
    x_{n+j}^{\mathrm{log}})$ for $n=12$ and summed
over all shell separations available
in the finite system.
The numerical results show excellent
agreement with the analytical theory,
including the discontinuous transition from
$\mathcal K=0$ in the inverse cascade to
positive values in the forward cascade.
Figure~\ref{figSM2}(c) illustrates the growth
of covariance amplitudes and the respective correlation
length in the inverse cascade as
$a\nearrow0$.

\begin{figure}[p]
\centering
\includegraphics[width=0.45\textwidth]{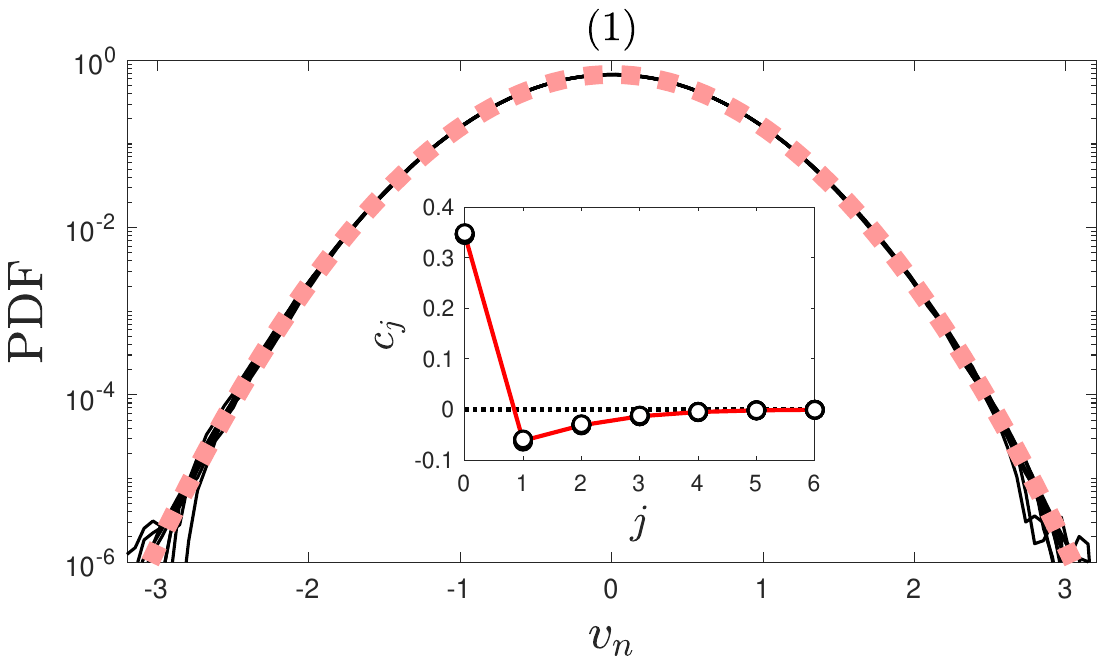}\\
\vspace{5mm}
\includegraphics[width=0.9\textwidth]{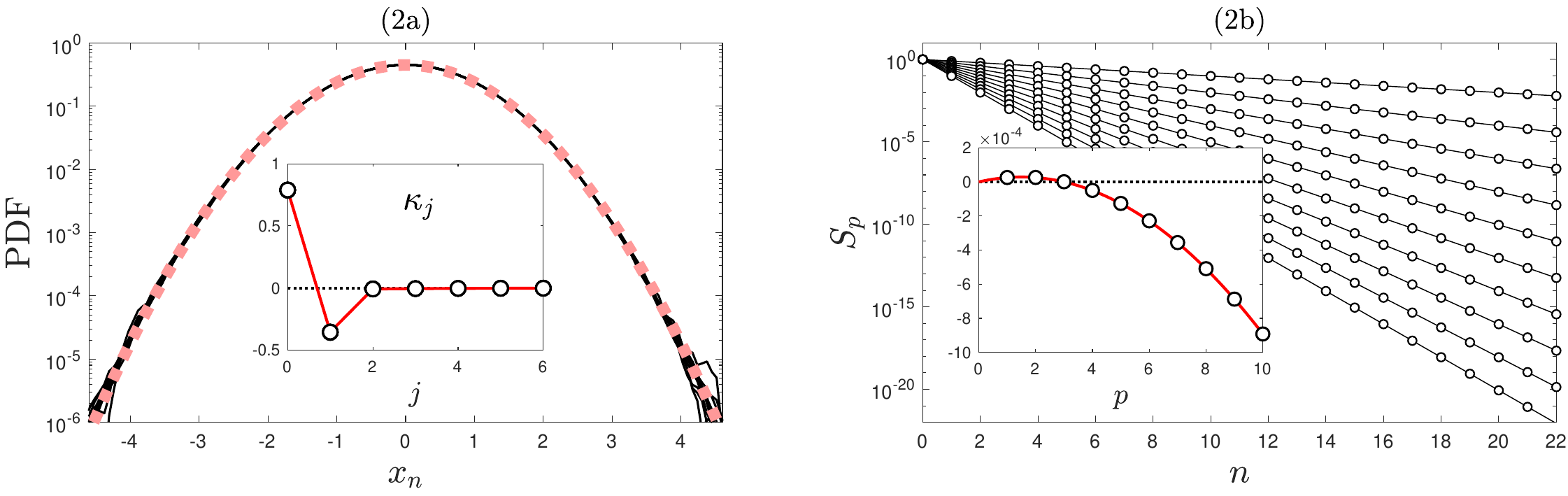}\\
\vspace{5mm}
\includegraphics[width=0.99\textwidth]{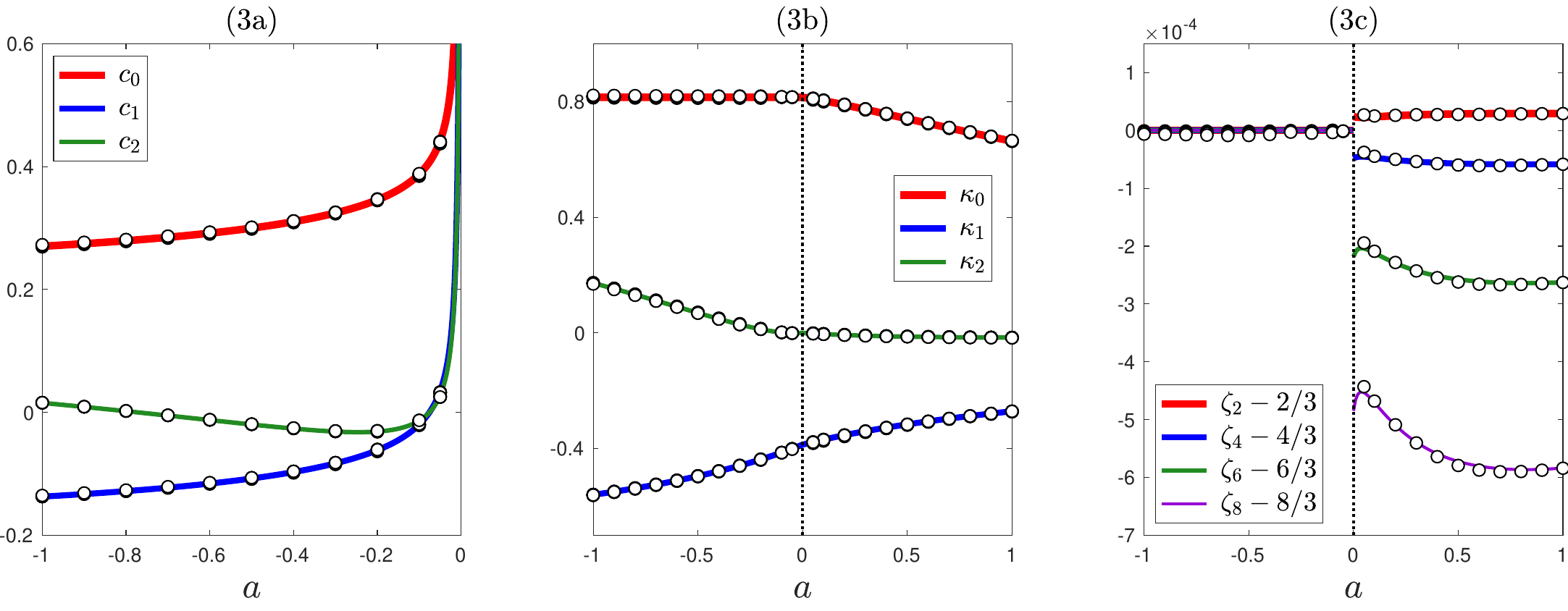}
\caption{
Additional simulation results for
$\varepsilon=0.02$, showing the same
observables as in the main Letter.
}
\label{figSM3}
\end{figure}

\begin{figure}[p]
\centering
\includegraphics[width=0.45\textwidth]{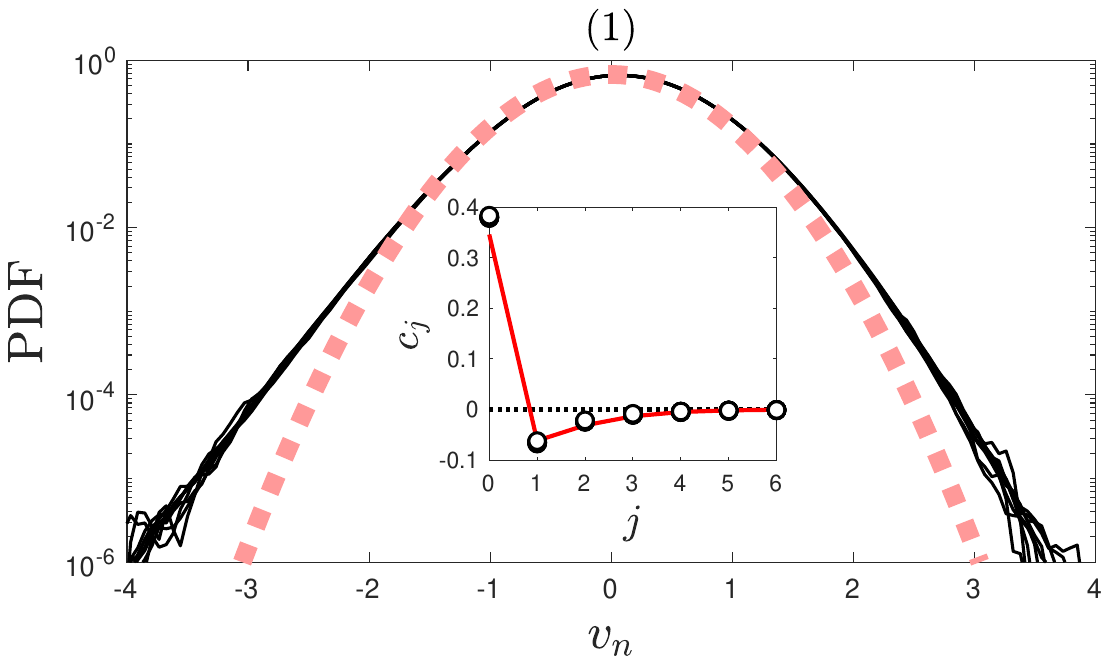}\\
\vspace{5mm}
\includegraphics[width=0.9\textwidth]{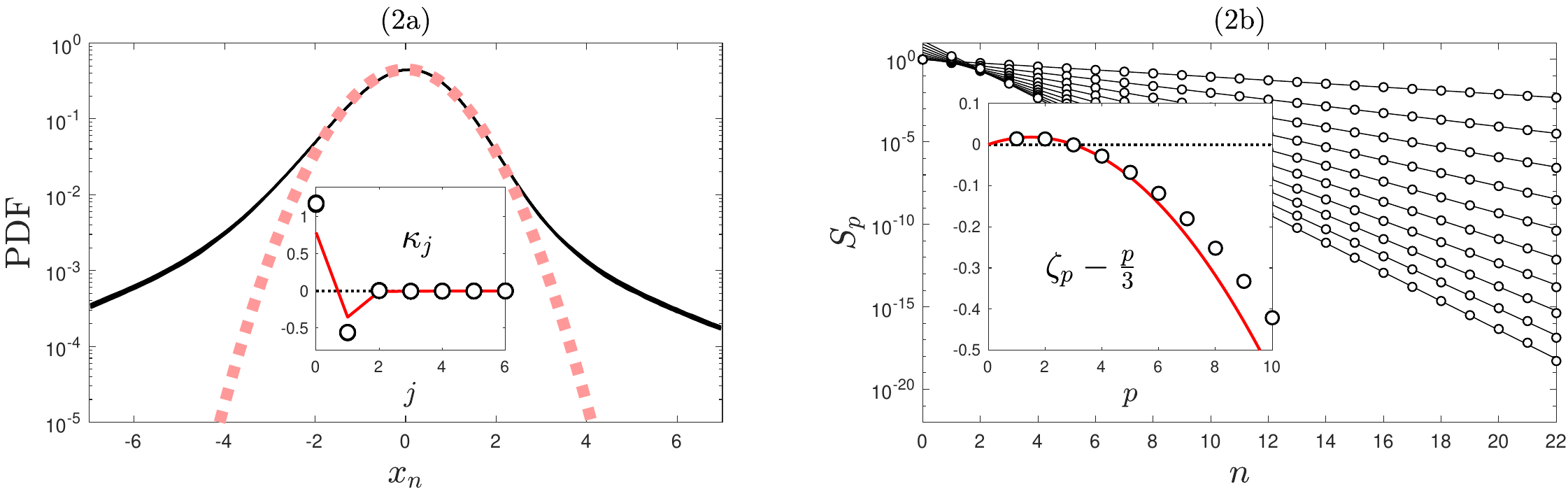}\\
\vspace{5mm}
\includegraphics[width=0.98\textwidth]{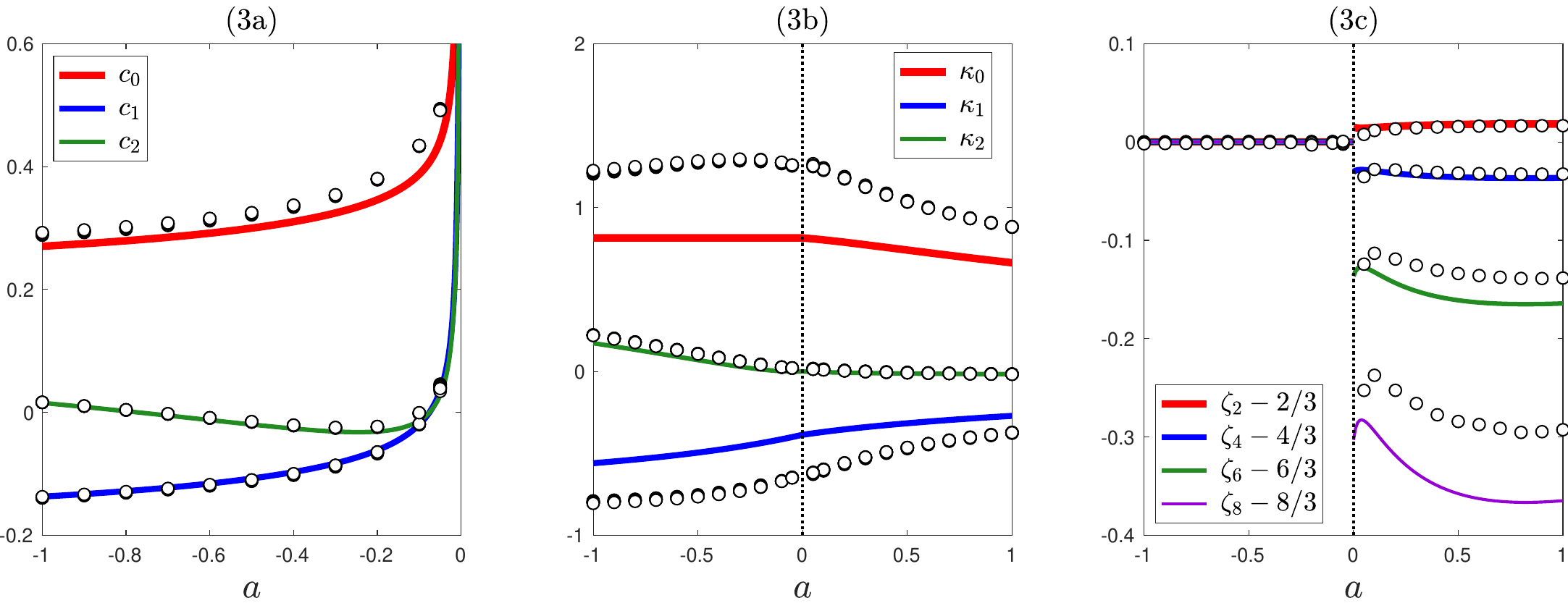}
\caption{
Same as Fig.~\ref{figSM3},
but for $\varepsilon=0.5$.
}
\label{figSM4}
\end{figure}

Figures~\ref{figSM3} and \ref{figSM4}
present corresponding results for weaker
($\varepsilon=0.02$) and stronger
($\varepsilon=0.5$) stochastic fluctuations.
For $\varepsilon=0.02$, the numerical
statistics show closer agreement with
the leading-order Gaussian theory,
whereas noticeable deviations from
Gaussianity appear at $\varepsilon=0.5$.
Nevertheless, the simulations retain the
main qualitative features of the cascade
transition: the growth of velocity
covariances in the inverse cascade near
$a=0$, the statistical self-similarity of
multipliers, and the sharp onset of anomalous
scaling in the forward cascade.
These results suggest that the statistical
distinction between the two cascade regimes
persists well beyond the weak-fluctuation limit.

\end{document}